\documentclass[conference]{IEEEtran}
\renewenvironment{quote}
               {\list{}{\rightmargin\leftmargin}%
                \item\relax\small\textquotedblleft\ignorespaces}
               {\unskip\unskip\textquotedblright\endlist}
\def\CU{\hbox{$\mathcal{C}\kern-.2ex\scriptstyle\cup$}}
\usepackage{cite}

\ifCLASSINFOpdf
\usepackage[pdftex]{graphicx}
\else
\fi
\usepackage[cmex10]{amsmath}
\usepackage{fixltx2e}
\usepackage{url}

\usepackage{amssymb}

\usepackage{pbox}
\usepackage{epigraph}

\begin{document}
%
\title{Behavior, Organization, Substance: Three Gestalts of General Systems Theory}

\author{\IEEEauthorblockN{Vincenzo De Florio}
\IEEEauthorblockA{PATS research group,
University of Antwerp \& iMinds Research Institute\\
Middelheimlaan 1, 2020 Antwerpen, Belgium\\
Email: https://www.uantwerp.be/en/staff/vincenzo-deflorio}
}


%



\maketitle

\begin{abstract}
The term gestalt, when used in the context of general systems theory, assumes the value of ``systemic touchstone'', namely
a figure of reference useful to categorize the properties or qualities of a set of systems. Typical gestalts used, e.g., in biology, are those
based on anatomical or physiological characteristics, which correspond respectively to architectural and organizational design choices
in natural and artificial systems.
In this paper we discuss three gestalts of general systems theory: behavior, organization, and substance, which refer respectively
to the works of Wiener, Boulding, and Leibniz. Our major focus here is the system introduced by the latter.
Through a discussion of some of the elements of the Leibnitian System, and by means of several novel
interpretations of those elements
in terms of today's computer science, we highlight the debt that contemporary research still has with this Giant among
the giant scholars of the past.
\end{abstract}



%
\IEEEpeerreviewmaketitle

\section{Introduction}
\epigraph{General Systems Theory [\dots] hopes to develop something like a ``spectrum'' of theories---a system of systems which may perform the function of a
``gestalt'' in theoretical construction. Such ``gestalts'' in special fields have been of great value in directing research towards the gaps which they reveal.}
{\textit{General Systems Theory---The Skeleton of Science}\\ \textsc{K. Boulding}}

The notion of a General Systems Theory is no recent invention. Already Aristotle proposed a tentative classification of ``systems''.
A common aspect between Aristotle's and all the classifications that followed is the use of one or more \emph{systemic touchstones},
namely privileged aspects that provide the classifier with ``scales'' to diversify systems along one or more dimensions.
A common term used to refer to such touchstones is 
gestalt\footnote{German: ``Essence or shape of an entity's complete form''~\cite{Jacks10}.}~\cite{Bou56}.

Not only the Great One started the discussion, but he also set most of its subsequent ``rules'' by classifying
systems according to several anatomical (that is, architectural) and physiological (organizational) gestalts.
Remarkably enough, Aristotle realized that a discussion purely based on the above aspects would not be
complete, and suggested to make use as gestalts also of behavior, purpose, and teleology---the very same
touchstones at the core of the renowned article~\cite{RWB43} by Rosenblueth, Wiener, and Bigelow.
He was also the first to put the accent on social behaviors by writing about mutualistic relationships between individuals~\cite{AnimBeha}.
This second type of gestalts put their privileged focus on the characteristics and the quality of the
\emph{emerging products\/} of systems rather than on their structural, i.e., constitutive, peculiarities;
thus on dynamic aspects rather than on static design choices. Quality in particular is expressed as
the result of a match with a deployment environment, which in turn may be assumed to be static or vary
with time.

As postulates in a geometry or the axioms in a conceptual system,
gestalts define the way we address a given problem and set the boundaries of what we can prove in it.
Furthermore, as already mentioned,
they have
``\emph{great value in directing research towards the gaps which they reveal}''~\cite{Bou56}.

The present contribution exemplifies
three well-known general systems theories: the behavioral system of Wiener et al.~\cite{RWB43} (Sect.~\ref{s:rwb});
the behavioral-organizational categorization of Boulding~\cite{Bou56} (Sect.~\ref{s:bou}); and, in Sect.~\ref{s:L},
the system of Leibniz,
based on the behavioral-architectural-organizational gestalt he refers to as ``substance''.
In each section we briefly discuss the system classifications stemming from the adopted gestalt
and highlight the research directions that they revealed.
In particular in Sect.~\ref{s:L} we provide a number of
modern-day interpretations of the major concepts in the philosophy of Leibniz.

Our conclusions are drawn in Sect.~\ref{s:end}, where
we highlight how the system of Leibniz anticipated several of the research directions that emerged
with the birth of the computer era, to the point that modern computer science time and again
provides a useful interpretation of the concepts found in Leibniz's philosophy.
In particular, we reflect on recent results such as the ones in companion paper~\cite{DF14b}
and point out our personal debt with the Leibnitian System.


\section{Behavior}\label{s:rwb}
In their renowned paper~\cite{RWB43} Rosenblueth, Wiener, and Bigelow introduce the concept of 
the ``behavioristic study of natural events'' and propose a classification of
systems that focuses on the
``change produced in the surroundings by the object''---namely the system's behavior.


The Authors' starting point is given by the classes of passive and active behavior.
They describe passive behavior as the one in which
``the object is not a source of energy; all the energy in the output can be
traced to the immediate input (e.g., the throwing of an object)''.
All behavior that is not passive is active, namely behavior
in which ``the object is the source of the output energy involved in a given [change]''.
The class of active behavior can be refined into two subclasses:
purposeful or non-purposeful active behavior.
The first subclass identifies systems that aim at achieving some goal, while the second
one characterizes ``random'' behaviors---behaviors that is exercised by systems that
are a source of change but whose action does not serve an apparent purpose.
In the latter category we may have for instance a source of radiations.

The class of purposeful behavior is then decomposed into two other subclasses:
teleological and non-teleological behavior, the first being characterized by
the presence of a feedback loop by means of which the system can continuously
adjust its action with respect to the intended purpose.
Non-teleological
behavior is the one in which said feedback loop is absent.

In the companion paper~\cite{DF14b},
systems
capable of teleological behavior have been described as \emph{reactive\/} systems.
Obviously as a prerequisite of reactiveness those systems are
\emph{open}~\cite{Hey98}
(namely able to perceive, communicate, and interact with external
systems and the environment.)

Finally, the Authors differentiate teleological behavior into yet another
couple of sub-classes: extrapolatory and non-extrapolatory behaviors.
In the former case the system is capable of advanced \emph{apperception},
which we defined in~\cite{DF12a}
as ``the ability to construct
theories about the current and related past situations with which to drive system evolution''.
In practice extrapolatory behaviors are those in which the feedback loop
is governed by the hypothesized future state of the goal---for instance,
its position. Moreover, in extrapolatory behaviors the hypothesis is drawn on the basis of
one or more context figures.
Extrapolatory systems are thus not merely able
to perceive the environment they are deployed in---they are also able
to store in some form the perception data; continuously correlate 
past and new data; create a model to predict the future state;
and use that model to steer the action of their feedback loop.
Extrapolatory behavior is thus \emph{proactive}~\cite{DF14b}
and corresponds to the so-called MAPE-K loop-systems of
autonomic computing~\cite{KeCh:2003}.
The Authors call the number of context figures used in
the predictive model ``the order'' of the behavior, which
constitutes a final sub-classification in their treatise.

\subsection{Conclusions}\label{s:rwb:c}
As observed by the Authors, a major consequence of the behavioristic approach is given by the fact
that it
does ``omit the specific structure and the intrinsic organization'' of the systems under scrutiny
and only focuses on the action produced by the system.
The model proposed by Wiener et al. thus does not concern itself with the nature of the system
or its design: in fact it may be applied to any ``object'' (the Authors' term for
``system''), be it natural or artificial, hardware or software, individual or collective, or
any mixture thereof. The only important figure in their discussion is the observed behavior,
namely ``the examination of the output of the object and of the relations of this output to the input.''

The behavioral gestalt, for the first time applied also to artificial
entities, provides researcher with a powerful tool to reason about
the quality of systems anticipating in particular results such as
autonomic computing, dependability, and resilience.
The significance of the work of Wiener in the 21st Century may be
also exemplified through recent works as the
companion paper~\cite{DF14b}, which proposes a behavioral interpretation
of the concept of system-environment fit and suggests its use
to let systems manage their own resilience provisions.

\section{Organization}\label{s:bou}
A similar approach to Wiener's is followed by Kenneth Boulding, who in addition to behavioral
features also focuses his attention on the organizational characteristics
of systems both natural and artificial~\cite{Bou56}. Boulding suggests an
``arrangement of \emph{levels\/} of theoretical discourse,'' which 
he names after 
systems best-representing each level: ``Thermostat'', ``Cell'', ``Plant'', ``Animal'', ``Human Being'', and others.

As already mentioned, the accent in Boulding is not only behavioral; this makes it
possible to highlight in particular
aspects such as the openness of the system~\cite{Hey98}; its ability to
be not just aware but also self-aware; as well as the ability to enact
collective forms of behaviors.

An important addition in Boulding's system with respect to Wiener's
is given by the new class of ``social organization'', namely systems composed by
``a set of roles tied together with channels of communication''. The new class corresponds
to social behavior, which may in fact be the subject of a classification of its own\footnote{More
	information on this is available, for instance, through~\cite{2014arXiv1401.5607D}.}.
Rather than a separate class, social behaviors may be interpreted as an attribute of the
behavioral classes of~\cite{RWB43} and in fact could be used as
a second ``coordinate'' for a general classification of behaviors.

Boulding also introduces a final class, consisting of hypothetical systems whose
organization and behaviors are beyond those of the class of Human Beings.
Such ``Transcendental systems'' are useful for a discussion of the quality
of systems as they represent a reference point as exemplified, e.g., 
in~\cite{DF12a}.

\subsection{Conclusions}
As mentioned already, Boulding's gestalt incorporates and extends Wiener's, thus several
of the considerations we stated in Sect.~\ref{s:rwb:c} apply here too.
An important additional research direction naturally stemming from the Boulding system is
the dense contemporary ``corpus'' of research that focuses on social organizations, social systems,
and social behaviors, including human and machine ecological aspects.

%
%
%

\section{Substance}\label{s:L}

\epigraph{Streets that follow like a tedious argument\\
Of insidious intent\\
To lead you to an overwhelming question.\\
Oh, do not ask, ``What is \emph{it}?''\\
Let us go and make our visit.}%
{\textit{The Love Song of J. Alfred Prufrock}\\ \textsc{T.S. Eliot}}



A different and more direct
approach is the one proposed by Leibniz and anticipated, to a much lesser extent, by Aristotle and Pythagoras.
In both Aristotle and Leibniz, the accent is also put on behaviors but more so on the systems producing them: the \emph{substance}.
Aristotle calls substance ``a subject that underwent change'', which is  a definition surprisingly
similar to that of Rosenblueth, Wiener, and Bigelow. 
And Aristotle too distinguishes passive and active-behaviored substances and calls the latter as \emph{entelechies}:
substances that ``bring about their own changes from one state to another''. As recalled in Sect.~\ref{s:rwb}, this is
in fact the same initial step taken by Wiener et al. when laying the foundation of their
behavioral method.
Leibniz makes this concept and term his own and also refers to substance as 
entelechy, namely
``a source of actions, or rather, its own actions''~\cite{leibniz2006shorter}; but he
introduces several novel
ideas\footnote{This plurality of ideas is reflected in a plurality
	       of terms to refer to substances, that Leibniz calls
	       minds, souls, entelechies, and monads depending on the aspects he wanted to highlight.}.
So many and intertwined are those
ideas\footnote{Strickland very eloquently
	       refers to Leibniz's as to a ``piecemeal approach to the diffusion of his ideas''~\cite{leibniz2006shorter}.}
that it is rather difficult
to expose them in a satisfactory unitarian way. In what follows we will not attempt such a titanic
task but rather will try to build a concise model of a subset of those aspects that best match 
the themes of the present contribution.

\subsection{Substances as Interconnected Networks}\label{s:L:nw}
In Leibniz, substances are fully interconnected networks of all-open, all-aware active-behaviored ``nodes''
(viz., entelechies)
whose behaviors depend deterministically on the influence exercised by all other nodes.
The term ``influence'' refers here to a general Law, called by Leibniz the \emph{Principle of Concomitance},
stating that a pre-established \emph{harmony\/} exists among all substances.
Though this harmony among substances is perfect, substances perceive other substances
depending on some finite quality called ``clarity of perception'' (or representation)---sort of
a metric function measuring a ``perception distance'' between any two substances.

Depending on the clarity of perception substances may influence each other differently. When the
influence is very strong the involved substances are said to ``embody''. Embodiment means that a set of substances
are so much mutually in perceptive relation with one another (so much ``in harmony,'' so to say) that they
give raise to a new, \emph{social\/} substance.
The social substance is represented by a controlling substance, which Leibniz calls ``Mind''. 
When substances are embodied the mutual influence is so strong that, e.g., stimuli travel quickly
from one region of the network to the other, thus creating a feeling of concomitance
for perceived events and sensations---such as the feeling of pain.
To set this concept with the language of modern technology we may think of a sensor network whose nodes
may directly interact only with the nodes in wireless reach.

In Leibniz, bodies, minds, and perception---including the perception of physical matter---are actually a
product of the above concomitance.
The principle is valid for all substances whatever their nature,
but the exercised influence may vary and be felt differently depending on the degree of clarity of
the involved substances. As an example,
the energy released by a star in a far galaxy or a butterfly flapping its wings a continent away may be so ``distant''
from us as to exercise a minimal influence on us---and in fact to go undetected at all.
At the other extreme of the ``clarity spectrum'', an offence experienced by a vital organ would be immediately
perceived by the social substance---in particular, by the Mind---and have a profound effect on the whole
social substance\footnote{Obviously the Leibnitian concept of a degree of clarity has little to do with
	the familiar notion of geometrical distance.}.

We observe how the above definition \emph{extends\/} considerably that of Boulding's social organizations
that we recalled in Sect.~\ref{s:bou},
his concept of degree of clarity basically corresponding to that of the communication channels in Boulding.
In fact Leibniz goes much beyond Boulding and even appears to anticipate (of about three centuries!)
several of the ideas of social
constructivism and in particular those of Actor-Network Theory~\cite{Latour06}.

The clarity of perception characterizing a substance is not an absolute and eternal property; rather, it has a finite span after which
the network---the social substance---disintegrates into its constituent substances: the substance 
``dies\footnote{Think again of the nodes of a sensor network deployed in unmanned territory and running on batteries;
		if batteries discharge beyond the possibility to transmit, the ``social substance'' collapses into
		a set of individual nodes.}''.
Moreover, Leibniz observes how the behaviors of a substance in close relation
with a second one may result in centrifugal forces that distort or dissipate either or both of the substances'
``bodies\footnote{Crosstalk or adjacent-channel interference may be used to exemplify this concept.}''.

\subsection{Substances as All-aware, All-open Systems}\label{s:L:aware}
As already mentioned, in Leibniz
changes ``ripple away'' from an originating substance (namely, from the active behaviors of an entelechy)
and are perceived by all others, albeit with different effects depending on their mutual ``distance''.
Being a general law of all substances, the Principle of Concomitance implies for Leibniz that
all substances must be ready to encode through some internal representation any of the possible events
occurring outside of them.
Leibniz's conclusions are that substances must be embedded with a mechanism to represent and instantly reflect
all the possible states of all the substances in the whole universe. This includes any change of state
due to ``rippling''.
Leibniz imagines also that this internal representation-and-reflection (RR) mechanism constitutes the only method of interaction
between substances. Substances are in fact ``a world apart'', as he states. With the terminology of computer science,
we could say that Leibniz imagines that substances run in separate ``process spaces'' and that their
all-awareness and their RR mechanism provide an indirect method of interaction based on
an internal representation of the ripples.

It is worth highlighting how the idea of an internal model, or representation, of the external world,
which of course is very much influenced from Plato's Cave,
closely corresponds to the modern concept of \emph{qualia\/} as introduced, e.g., in~\cite{Kanai12} and
discussed in~\cite{DF12a,DF13b}\footnote{In particular, from~\cite{DF13b}: ``sensors [..]
	reflect a given subset of the worlds raw facts into internal representations that are
	then stored in some form within the systems processing and control units---its ``brains''.
	Qualia is the name used in literature to refer to such representations.''}.

In fact, Leibniz asserts, even the production of a new qualia state produces a change; and that change also ``ripples'',
as any other behavior, from the originating substance to all others, leaving a footprint that is proportional
to the degree of clarity between substances. I like to refer to the overall effect
of these reflections and reactive behaviors as to a gigantic ``metaphysical storm.''\footnote{For the Reader
	accustomed to the Twitter social system a way to represent such ``storm'' would be that of considering
	a circle of users that consistently re-tweets any message received by the members of the circle---including
	re-tweet notifications!}



\subsection{Substances as Conceptual Models}\label{s:L:mdl}
We said already that in Leibniz substance is a unicity.
He adds that substances are unicities that produce actions ``in accordance with their own individual concepts''.
What makes each and every substance unique and different from all others is
indeed \emph{the concept\/} of that substance: its \textbf{identity}, which makes it \emph{in-dividual}, namely
conceptually non-divisible.
In other words a substance is an entity whose concept is so peculiar and so \emph{strong\/} as to shift the attention
from its parts to an emerging unity---from the components to the composed. ``Man,'' for instance, is a substance,
because it is characterized by a concept
that is so complete and well-defined that we do not see the complex hierarchies of sub-systems a man consists of; rather, we just see
the product \emph{emerging\/} from the interactions---the social behaviors and systemic features we could say---of those sub-systems.
Once more we can highlight here the strong link with the philosophy of Pythagoras~\cite{Iamblichus}.
Making use of an American vernacular we could say that the substance of Man is what makes him \emph{thick}.
In fact, this is precisely what Leibniz asserts:
substances are
the \emph{only\/} actual form of existence, while the so-called physical world is nothing but
a distorted perception due to a limited ``power of representation''---an argument that
clearly reminds of Plato's Cave. The only reality is in fact
that of substances, and substances are
\emph{conceptual models}, namely
system templates. One such substance is, for instance, the algorithm of Bubble Sort:
a conceptual unity that results from a network of ancillary substances in harmony
with one another and emerging as a unequivocally identified substance different from all possible others.
The quality of Bubble Sort is that,
quoting Aristotle, it is ``more than the parts it is made of''. The network of ancillary concepts that
constitute Bubble Sort produces a peculiar added value, a purposeful behavior that results in
a method to sort objects. Thus, Leibniz tells us,
Bubble Sort is characterized by ``a certain demand for existence''~\cite{leibniz2006shorter}: it ``deserves'' to exist.
Of course other substances exist whose collective emerging behavior results in a similar service.
Quick Sort is indeed another such substance, and it is also characterized by its own
``claim to existence''~\cite{leibniz2006shorter}. Depending on the ``systemic quality'' of similar substances,
some of them are
``conceived\footnote{From \emph{concipere}, whose meanings include ``to become pregnant''\cite{w:conceive}.}''
by ``God,'' namely selected for existence,
while some others are discarded---for instance due either to limitations or to some ``natural'' tendency towards elegance and
conciseness.
Another reason for the selection of a substance is given by the fact that ``not all possible substances are
\emph{com}possible''~\cite{leibniz2006shorter}, viz. mutually compatible.
Two examples of this compossibility come to mind:
\begin{itemize}
	\item An ``Ultimate Predator'' substance would prohibit the existence of other ``prey substances'' and eventually
result in its own demise---as can be inferred from the Lotka-Volterra equations. Thus nature---or, for Leibniz, ``God''---prevents
such a ``compossibility'' to occur.
\item The axioms in geometry $E$ and the theorems that one can demonstrate in it are compossible concepts in $E$, but may well
	be that certain concepts that are ``valid'' in $E$ may contradict the concepts in another geometry $\neg E$; thus
	they would be not compossible in $\neg E$.
\end{itemize}
	
Remarkably enough, we can observe once more how the above concept of a ``systemic quality'' introduces a
classification:
\begin{itemize}
	\item Certain substances, e.g., Bubble Sort or Quick Sort, exhibit no form of awareness.
		In other words, they construct no model whatsoever of themselves or their environment.
		They correspond to Wiener's servo-mechanisms and are only capable of purposeful behaviors.
	\item Other substances, such as Cells and Plants, are characterized by primitive and
		very limited forms of awareness and ``openness''~\cite{Hey98}.
		They are only able to construct a very limited model of their ``world'' and strive
		towards basic teleological goals---for instance, survival.
	\item Yet others, such as Animals, have primitive forms of self-awareness.
		Their model of the physical reality is more complex and translates in
		simple proactive behaviors. A limited model of the ``self'' is also
		under their grasp.
	\item Substance Man reaches an even greater ability to exert complex behaviors and
		reach high degrees of self-awareness and consciousness. Man in fact is
		even able to reason about the nature of substances
		and construct theories---such as Leibniz's---about the working of the ultimate
		``network of networks''---the universe.
\end{itemize}

As one can clearly realize, this results in a 
general systemic classification---a general systems theory---not dissimilar from Boulding's and Wiener's.
Substances are characterized by different ``fidelity'', which we defined in\cite{DF14a}
as ``the compliance between corresponding figures of interest in two separate but
communicating domains''. These two domains in Leibniz are actually the Qualia world and the Physical world,
the former being the result of the RR mechanism introduced in~\ref{s:L:aware} while the latter
is the metaphysical reality---what Leibniz considered to be ``the Mind of God''. It is
there that conceptual models are conceived and it is from there that they are ``set in motion.''

%
We observe how the above concept of substance as a model is in fact very much intertwined with that of its uniqueness
and identity. 
Substance is a peculiar and well-defined ``logic'' that is different from all others---as in Aristotle's
concept of definition. Aristotelian entelechy is in fact also the ability to retain this
identity\footnote{Sachs~\cite{Sachs} translates entelechy as ``being-at-work'' while ``staying-the-same'':
		  ``a source of actions, or rather its own actions''~\cite{leibniz2006shorter}
		  that strives to retain its identity---namely its peculiar conceptual foundations.}.
As already mentioned, any substance, e.g. Bubble Sort, is a concept that is itself and no other one---a unique
concept in other words, in that modifying it even slightly would turn it into something else---a variant.

Leibniz dreamed of a knowledge representation language in which any conceptual model---any substance---would have
been expressed, and of a tool to verify ``mechanically'' the validity of predicates expressed in
that language and requiring evaluation (so-called contingent truths). He called 
\emph{Characteristica Universalis\/} (\CU) the language and \emph{Calculus Ratiocinator\/} the tool.
\CU{} was a diagrammatic language employing pictograms. The pictograms were convenient representations
of modular knowledge of any scale, with segments representing different properties---for instance
compossibility or non-compossibility. Leibniz exemplified this with well-known diagrams such as the one on the
frontispiece of his \emph{De Arte Combinatorica}.
There the four basic Aristotelian components are depicted together with lines stating whether
any couple of components would be compossible or otherwise, as well as which properties would emerge from their
union.

The \CU{} language is Leibniz's way to represent substances as networks of other substances,
together with their relationships.
Pictograms represent modules, namely knowledge components packaging other ancillary knowledge components.
In other words, pictograms are Leibniz's equivalent of Lovelace's and Turing's tables of instructions;
of subroutines in programming languages; of boxes in a flowchart; of components in component-based software
engineering\footnote{It is also intriguing to observe how compossibility and non-compossibility correspond
	to the concept of component interfaces. Figure~\ref{f:ChUniv} exemplifies this with a component application
	for encrypted data communication.}.  
It is no surprise that Leibniz observed that ``mankind is still not mature enough to lay claim to the advantages
which this method could provide''~\cite{LeibnizStrickland} and that
``a far greater secret lies hidden in our understanding, of which these are but the shadows''~\cite{LeibnizLoemker}.

As we already mentioned,
compossibility and quality determine a substance's claim for existence.
But the evaluation of compossibility and especially quality calls for
matching the substance with external conditions---an environment.
This concept is strikingly in line with the methodological assumption in our
companion paper~\cite{DF14b}:
\begin{quote}
	Our starting point here is the conjecture that
	        [quality] \emph{is no absolute figure}; rather, it is \emph{the result of a match
		with a deployment environment}.
\end{quote}

\begin{figure}
	\centerline{\includegraphics[width=0.15\textwidth]{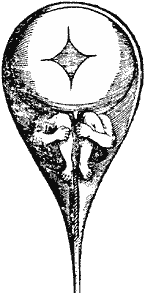}}
\caption{Preformationism exemplified by Nicolaas Hartsoecker, 1695. Image from the Wikimedia Commons.}\label{f:pref}
\end{figure}

\begin{figure}
	\centerline{\includegraphics[width=0.3737\textwidth]{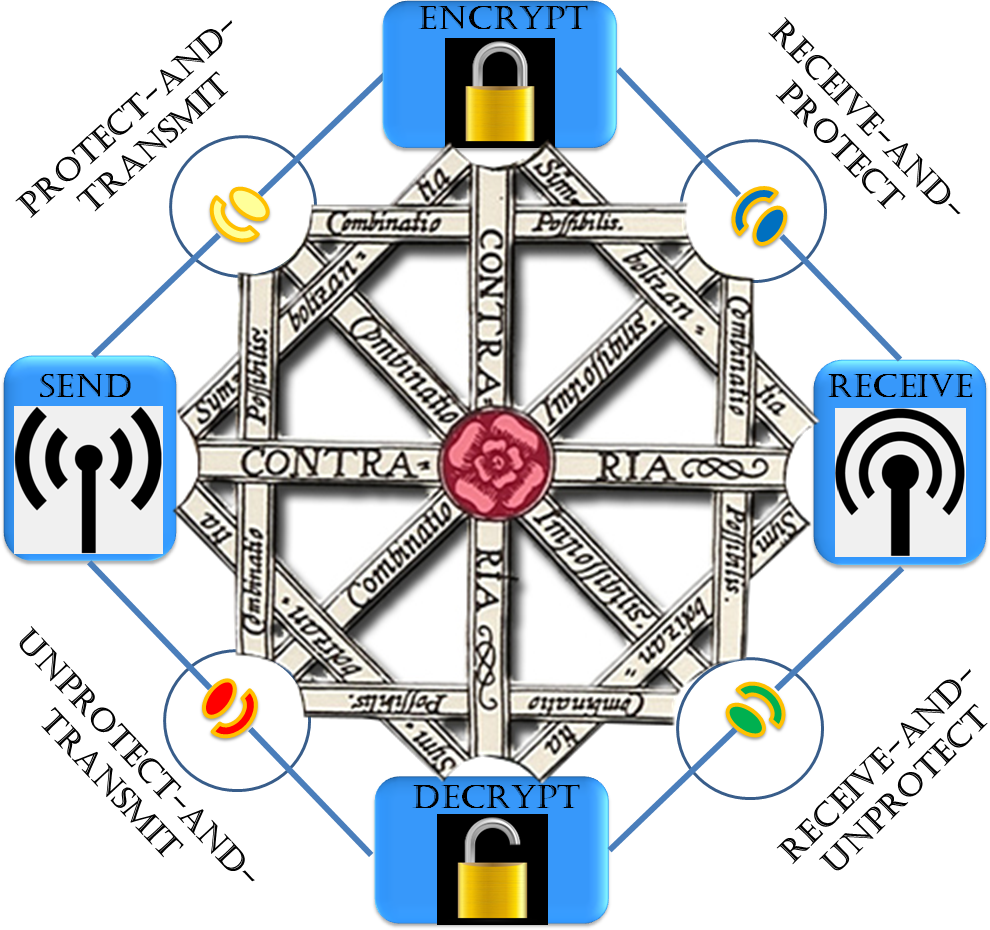}}
	\caption{The well-known diagram in the frontispiece of Leibniz's \emph{De Arte Combinatorica\/}
	reinterpreted in view of component-based software engineering.
	Interface color represents compatibility/compossibility: different colors imply 
	interface incompatibility (corresponding to pictogram incompatibility and substance non-compossibility).}\label{f:ChUniv}
\end{figure}

\begin{samepage}
Remarkably enough, Leibniz introduces the same methodological assumption. A fair selection of a coherent
set of compossible substances requires a complete assessment of the
quality of its constituents; but the only way to achieve such a complete assessment is
by confronting the substances with a vast amount of environmental conditions and checking
their individual and collective behaviors. The ``open variables'' in the substances---corresponding
to variables in \CU{} language ``scripts''---are then \emph{grounded\/} with respect to
various contextual conditions. This operation is called by Leibniz \emph{unpacking\/} and corresponds
to solving a logic expression by assigning ``facts'' (truth values) to its open variables until
the expression becomes either a tautology or a contradiction.
\nopagebreak
Substances are thus concepts, or better, ``scripts,'' expressed in \CU{} language\footnote{In fact Leibniz considers
			  substances as ``second-order'' scripts in that they are the product of
			  a first-order script, similarly to the spermatic animalcules theorized by Thonis 
			  van~Leeuwenhoek. Leeuwenhoek, also known as the Father of Microbiology,
			  was the first to observe spermatozoa. He was also
			  the grand developer of preformationism,
			  in turn derived from Pythagoras and Aristotle. Preformationism states that all beings
			  are the development of preformed miniature-versions---the above mentioned animalcules.
			  Figure~\ref{f:pref} exemplifies preformationism showing a homunculus within a spermatozoon.
			  Leibniz visited van Leeuwenhoek and was a convinced
			  believer of his theories, which he adopted in his own System.}.
\end{samepage}

The assessment is not just individual in that it is also applied, ``by construction'' so to say, to the
whole current set of compossibles---namely, to the whole current ecosystem of substances.
As mentioned already, through the so-called ``rippling'' assigned facts are propagated to all other substances
as in a sort of ``universal gossiping''~\cite{DFB12a}
among the nodes in a sensor network. Compossibles are confronted and selected also considering their
\emph{entelechy}, namely their ability to retain their conceptual identity~\cite{Sachs}---with the terminology of
modern computer science, their \emph{resilience}~\cite{DF13a}.

Another criterion for the mentioned ecosystem-wide assessment is given by the fact that the receptivity of the world
is limited and ``God'', namely ``a certain divine mathematics''~\cite{Leilines}, aims ``naturally'' at making the
best of the available resources.
The words of Leibniz are particularly remarkable:

\begin{quote}
	Out of the infinite combinations
	of possibles, and possible series, there exists one through which the
	\textbf{greatest amount of essence or possibility} is brought into existence. There is
	always in things a principle of determination which must be sought in
	maximum and minimum; namely, that the greatest effect should be
	produced with the least expenditure, so to speak. And here the time, the
	place, or in a word \textbf{the receptivity} or \textbf{capacity of the world}, can be
	considered as the expenditure or the land on which a building is to be
	constructed as fittingly as possible, while the variety of forms correspond to
	\textbf{the fitness} of the building and to the number and elegance of its rooms.
	And the situation is like that in \textbf{certain games where all the spaces on the
	board are to be filled according to certain rules}, and where, unless you use
	some skill, you will in the end be excluded from certain spaces and forced
	to leave more spaces empty than you could have or wished to. But there
	is a definite rule through which the maximum number of spaces is most
	easily filled. [\dots] In short it is just like tiles that are arranged so
	that as many as possible occupy a given area~\cite{leibniz2006shorter}.
\end{quote}

For this author it is remarkable how in such a relatively limited passage Leibniz condenses
so large a variety of concepts and ideas whose significance is particularly apparent
in our modern times. He discusses of limited receptivity---a
concept which reminds of the ideas expressed in the renowned ``Tragedy of the Commons'' paper~\cite{Hardin68};
of system-environment fit---the cornerstone of our discussion in our companion paper~\cite{DF14b};
and his vision of the world as a board game leads naturally to concepts such as cellular automata,
virtual reality, multi-agent systems,
and artificial life. Moreover, his criterion of reaching as great a variety as possible among substances
matches remarkably well the results discussed, e.g., in~\cite{Knoll,2014arXiv1401.5607D}, namely
the key role played by diversity and disparity in the survival of biological (and digital~\cite{Eraclios13.11.12}) ecosystems.

\subsection{The Substance Scheduler}

As mentioned already, Leibniz conjectures the existence of a transcendental entity---a ``God''.
As in Boulding, said entity represents the highest level in the gestalt hierarchy.
But while in Boulding this concept is left unexplored, in Leibniz it is justified through a series of
logic deductions that follow from the very postulates of his system.

The very first of such deductions is stated through the following famous quote:
\begin{center}
	\emph{All substances ``subsist in the mind of God''}~\cite{leibniz2006shorter}.
\end{center}
The elements so far introduced allow us to attempt a daring interpretation of the above sentence:
the Leibnitian God is yet another substance, namely a network of substances with a central organization
and a central ``hub'' that embodies
(we could say, ``punctualizes''~\cite{Latour06})
the whole network
into a unique and in-dividual concept (cf. Sect.~\ref{s:L:nw}). The Mind of such network is God,
and the matter emerging from its union is the world.

Stated in other words, God is the largest possible network of networks---the largest possible ``scale''
in a gigantic recursive structure that spans an entire theory of concepts.
One may possibly visualize this through the
image of an enormous  mind-map connecting, e.g., all the arithmetically
derivable concepts, and with a predefined Center representing the whole system---in this example, the
concept of Arithmetic. Furthermore,
in Leibniz it follows that God is the central controller of the universe; a substance so perfect as to have
the utmost clarity of representation of all the other substances whatever their scale (whatever their level of
recursive nesting, that is). The most perfect substance thus; but a substance nevertheless, hence a concept (cf.
Sect.~\ref{s:L:mdl}); hence, the executor of a ``function''. To Leibniz this function can only be
\textbf{Ultimate Sort}: a ``procedure'' for the optimal scheduling-for-existence of the
available concepts. Thus God is an \emph{ordinateur}, or an operating system if you like,
who manages a limited process space and selects process images to be deployed and executed onto
the Bare Machine. A task, says Leibniz, not dissimilar to that of a player of a board game
in which the goal is being able to ``maximize the returns'', namely the overall quality,
by choosing the best and allocating the most compossible substances to have ``on board''.
God\footnote{Here the outstanding question is obviously: is the God we are talking about the ultimate level
	or merely the next level in a metaphysical hierarchy? Is he actually ``God'' or merely a demiurge---a public agent? Is he the
	One or his middleware?}
is therefore a sorting algorithm and his data structures are the substances and the world---in particular
its intrinsic limitations and current state. Ultimate Sort is written in \CU{} language and
is to be executed on a compliant machine---the already mentioned ``Calculus Ratiocinator'',
interpreted here as an execution engine for \CU{} language scripts.

So logic and coherent is Leibniz's discussion that by considering the major elements of the Leibnitian system 
briefly summarized in this section
it is possible to formulate a (non-pictorial) pseudo-code for Ultimate Sort as stated in Table~\ref{t:US}.

\subsection{Conclusions}\label{s:mod}
Our major lessons learned from the above discussion may be summarized as follows:
substance is in Leibniz also a module---a concept-network packaging a quantum of knowledge that becomes a new ``digit'', a new
concept so unitary and indivisible as to admit a new pictorial representation, a new and unique monad-symbol.
\CU{} is the ``general algebra in which all truths of reason would be reduced to a kind of calculus''~\cite{LeibnizLoemker}.
The fundamental properties of such language are:
\begin{itemize}
	\item First, its being compositional and modular 
		by construction\footnote{For Leibniz \CU{} is the language of the ``true characteristic [of substances,]
		which would express the composition of concepts by the combination
		of signs representing their simple elements, such that the correspondence between composite ideas and their symbols
		would be natural and no longer conventional''~\cite{LeibnizCouturat}.}. In other words,
		it is an isomorphic language, such that \emph{concepts are preserved through their compositions}.
	\item 	Secondly, its ability to reflect God's ``greater secret,'' through which ``a fundamental knowledge
		of all things will be obtained''~\cite{LeibnizParkinson}. What is this secret, what is this fundamental knowledge?
		We conjecture this secret to be
		what Leibniz calls \emph{art of complication}: ``when the tables of categories of our art of complication
		have been formed, something greater will emerge'' (ibid.) In other words, God's  greater secret is the
		ability intrinsic in nature's
		``divine mathematics'' to construct ``naturally'' ever more complex structures, ever more \emph{evolved\/}
		substances~\cite{WaAl1996}.
\end{itemize}

\begin{table}
\begin{tabbing}
\hspace*{3ex}\=\hspace*{2ex}\=\hspace*{2ex}\=\hspace*{2ex}\=\hspace*{2ex}\=\hspace*{2ex}\=\hspace*{2ex}\=\hspace*{2ex}\kill
\textbf{Procedure USort} ($S$, $W$)\\
/* $S$ is the set of all substances */\\
/* $W$ is a variable reflecting~\cite{DB07a} the state of the current world */\\
\textbf{begin}\\
\textsf{01}\>\textbf{Parallel Do}\\
\textsf{02}\>\>\textbf{At Individual Level}\\
\textsf{03}\>\>\>\textbf{For All $s\in{S}$ Do}\\
\>\>\>\> /* IntrinsicQuality returns the static component of substance quality */\\
\>\>\>\> /* This may include, e.g., the behavioral class of $s$ (cf. Sect.~\ref{s:rwb}), */\\
\>\>\>\> /* or the Boulding level of $s$ (cf. Sect.~\ref{s:bou}), */\\
\>\>\>\> /* or other architectural/organizational/behavioral characteristics. */\\
\textsf{04}\>\>\>\> $i\leftarrow$ IntrinsicQuality($s$);\\
\\
\>\>\>\> /* ExtrinsicQuality returns the dynamic component of substance quality */\\
\>\>\>\> /* It calls Unpack($s$, $W$) to execute $s$ with environmental conditions as in $W\kern-.1em.$ */\\
\>\>\>\> /* Variables of $s$ requiring the truth value of contingent truths are thus resolved. */\\
\textsf{05}\>\>\>\> $e\leftarrow$ ExtrinsicQuality($s$, $W$);\\
\\
\>\>\>\> /* IndividualQuality returns a substance's overall quality */\\
\>\>\>\> /* this corresponds to the concept of system-environment fit */\\
\>\>\>\> /* as defined in the companion paper~\cite{DF14b}. */\\
\textsf{06}\>\>\>\> $s.q \leftarrow$ IndividualQuality($i, e$);\\
\textsf{07}\>\>\>\textbf{End For}\\
\textsf{08}\>\>\textbf{End Level}\\
\textsf{09}\>\>\textbf{At Social Level}\\
\>\>\> /* If $W$ allows another substance to be deployed\ldots */\\
\textsf{10}\>\>\> \textbf{if} Receptivity$(W\kern-.1em)$ $>0$ \textbf{then}\\
\>\>\>\> /* \ldots a compossible substance of highest individual quality is selected\ldots */\\
\textsf{11}\>\>\>\> $s\leftarrow$ SelectForExistence$(S)$;\\
\\
\>\>\>\> /* \ldots and deployed in $W\kern-.15em.$ */\\
\textsf{12}\>\>\>\> Deploy$(s)$;\\
\textsf{13}\>\>\> \textbf{else}\\
\>\>\>\> /* If $W$ has reached its limits then we need to make room. */\\
\>\>\>\> /* We select the substance that has the ``worst-quality'' with respect to */\\
\>\>\>\> /* IndividualQuality and world-specific criteria (minimal loss of */\\
\>\>\>\> /* diversity and disparity~\cite{Eraclios13.11.12,2014arXiv1401.5607D}, maximal ``cost'' in terms */\\
\>\>\>\> /* of world space requirements, etc.) */\\
\textsf{14}\>\>\>\> $s\leftarrow$ SelectForDestruction$(S)$;\\
\\
\>\>\>\> /* The minimum-quality existing substance is purged. */\\
\textsf{15}\>\>\>\> Undeploy$(s)$;\\
\textsf{16}\>\>\> \textbf{end if}\\
\textsf{17}\>\>\textbf{End Level}\\
\textsf{18}\>\textbf{End Do}\\
\textbf{end.}
\end{tabbing}
\caption{Pseudo-code of Leibniz's Ultimate Sort.}\label{t:US}
\end{table}

\section{Conclusion}\label{s:end}
\epigraph{Whatever happens in a piece of music is nothing but the endless reshaping of a basic shape. Or, in other words, there is nothing in a piece of music but what comes from the theme, springs from it and can be traced back to it; to put it still more severely, nothing but the theme itself. Or, all the shapes appearing in a piece of music are foreseen in the ``theme.''}
{\textsc{Arnold Schoenberg}~\cite{Schoe1975}}

We have described elements of three general systems theories by focusing on their conceptual touchstones---their gestalts.
As anticipated by Boulding, each gestalt helps discuss a peculiar aspect of a family of systems
and ``directs research towards the gaps that it reveals''.
Particular attention has been devoted to \emph{substance}, the gestalt at the core of the Leibnitian treatise.
We have highlighted how several key ideas of modern science may find their foundation in the
system of Leibniz---including, e.g., virtual reality, artificial life, genetic programming,
autonomic computing, component-based software engineering, knowledge representation
languages, automatic deduction, cyber-physical things, and cyber-physical societies.
In particular we put the accent on knowledge representation and knowledge processing---namely
on the concept of \CU{} language and Calculus Ratiocinator engine---which Leibniz considered as the key
tools for ``that science in which are treated the forms or formulas of things in general, that is, quality in general''~\cite{LeibnizLoemker}---in
other words, General Systems Theory.

In this final section we like to acknowledge our personal debt with the system of Leibniz in two of our own recent research directions,

		A first example may be found in companion paper~\cite{DF14b}. In that work we introduce
		an intrinsic quality parameter given by the behavioral class of the system under scrutiny.
		This corresponds to Statement~\textsf{04} in Table~\ref{t:US}, in which the substance scheduler
		evaluates the intrinsic quality of a substance.
		We also define a system-environment fit---which corresponds to evaluating a substance's extrinsic quality,
		or the quality under specific external conditions. This is the same as in Statement~\textsf{05}
		in Table~\ref{t:US}: a behavioral implementation of function ``ExtrinsicQuality'' may be in fact
		that exemplified in Fig.~2 of the companion paper.

\begin{figure*}
	\centerline{\includegraphics[width=1.0\textwidth]{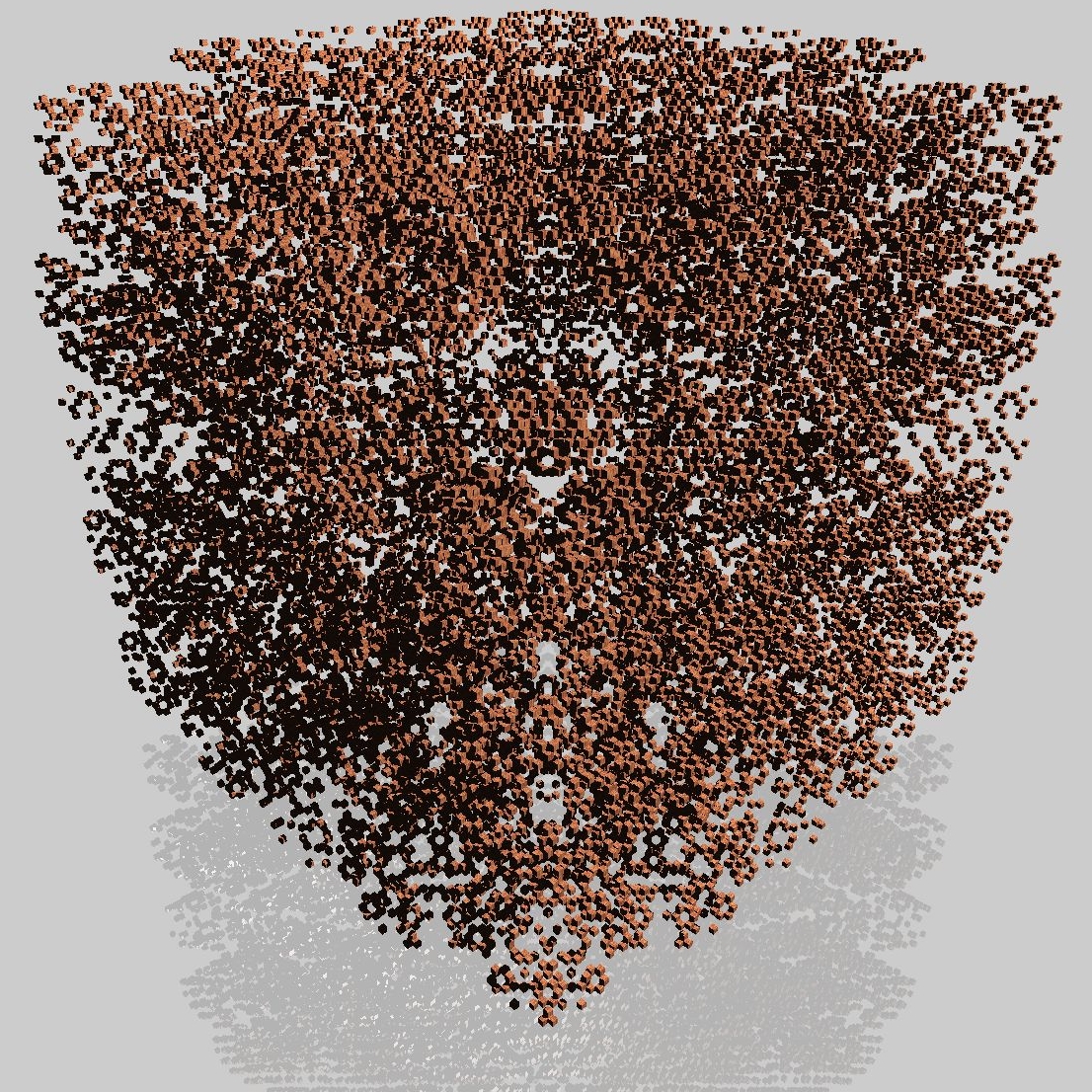}}
	\caption{Fractal social organization ``001123344''~\cite{DF13c} provides a geometrical interpretation
	of the Leibnitian concept of substance.
	The modularity emerging in the geometrical representation is a direct consequence
	of the modularity in the ``seed'': modules correspond in fact to sub-strings of ``001123344''.
	Drawing inspiration from the terminology of preformationism we could say that the seed is in this case
	a ``sociunculus'' of a society.}\label{f:fso}
\end{figure*}

	A second example is in the work presented in~\cite{DF13c} and anticipated by the mathematical models
		of the HeartQuake game~\cite{DeFl96} and of permutation numbers~\cite{DeFl05}. In those articles
		we considered the
		``movements'' produced respectively by deterministic game procedures; by the permutations of a fixed ``population''
		of digits; and by the non-deterministic arrangements of actants that respond to the onset of
		environmental conditions---such as natural or man-induced disasters. The graphs representing the collection of
		all possible arrangements are indeed networks of ``concepts'' embedding other concepts
		into recursive structures that, through some \emph{divine mathematics}, result in
		self-similar ``matryoshka doll'' graphs such as the one exemplified in Fig.~\ref{f:fso}.
		The modular structure in that figure is in fact the pictorial expansion of string ``001123344''---a
		5-pictogram representation of a substance if you like.
		As that string includes in itself a number of sub-strings, likewise its expansion
		includes the expansions of its sub-strings, with a conservation of modularity that reminds
		of the results in~\cite{WaAl1996}. This provides a geometrical interpretation of
		Leibniz's vision of the monads as networks of substances emerging and ``descending''
		from a central concept---what Schoenberg would probably refer to as a ``theme''~\cite{Schoe1975}
		a whole composition \emph{springs from\/} and may be \emph{traced back to}.
		An example of this principle is given by
		musical compositions such as \emph{Ostinato 011112333}~\cite{Ostinato011112333}---a
		musical rendition of the very same ``divine mathematics'' presented in Fig.~\ref{f:fso}
		but this time springing from substance ``011112333''.
		Every single note expressed in the mentioned composition derives in fact deterministically from 
		its ``theme'', string 011112333.

As a final remark we would like to draw once more the attention of the reader to 
van~Leeuwenhoek and his theory of preformationism---a theory that was enthusiastically accepted by Leibniz
and never doubted in the course of his whole life~\cite{leibniz2006shorter}. We conjecture that the
main reason for this may be that, though obviously an incorrect and unscientific concept, preformationism contains
\emph{in nuce\/} a quite modern and ``scientifically discussed'' concept, namely the already mentioned
principle of \emph{conservation of modularity}, viz. the property of conserving modularization when passing from
a genotypical representation (viz. a concept, namely an abstract and general template) to a phenotypical representation
(namely a particular ``realization'', or concrete expansion, of that template)~\cite{WaAl1996}. This property, which may
be probably best represented through the mathematical concept of an isomorphism between genotypical and phenotypical
algebraic domains, is in fact compatible with the Leibnitian vision of substances as ``second-order scripts'' produced by
``first-order scripts''.
As already mentioned, this conservation of modularity possibly hints at the ``greater secret'' hinted at by Leibniz,
namely the reasons why evolution ``evolves'', and
why nature ``naturally'' develops ever more complex \emph{substances}.
%

\section*{Acknowledgments} 
My gratitude goes to Professor Lloyd Strickland who kindly provided me with
comments and suggestions about this article.
Many thanks to Dr. Adriana Danielis (\url{https://twitter.com/AdrianaDanielis})
for the many and ever so insightful philosophical discussions.
My thanks go also to Dr. Tom Leckrone (\url{https://twitter.com/SemprePhi}),
Dr. Giovanni Fanfoni (\url{https://twitter.com/GiovanniFanfoni}), and
Dr. Stefano Rocca (\url{https://twitter.com/StefanoRocca59})
for sharing with me several works
about the philosophies of Pythagoras and Leibniz.

\bibliographystyle{IEEEtran}


\end{document}